\documentclass{aa}
\usepackage{graphicx}
\usepackage{txfonts}
\usepackage{natbib}
\def\cgs{erg\,cm$^{-2}$\,s$^{-1}$}
\begin{document}
   \title{RBS1423 - a new QSO with relativistic
    reflection from an ionised disk\thanks{Based on observations obtained
    with XMM-Newton, an ESA \mbox{science} mission with instruments and contributions
    directly funded by ESA Member States and NASA.}}

   \author{M. Krumpe
          \inst{1}
          \and
          G. Lamer
          \inst{1}
          \and
          A.D. Schwope
          \inst{1}
          \and
          B. Husemann
          \inst{1}
          }

   \offprints{M. Krumpe}

   \institute{Astrophysikalisches Institut Potsdam,
              An der Sternwarte 16,
              14482 Potsdam, Germany\\
              \email{mkrumpe@aip.de}
              }
   \date{Received 28 December 2006; accepted 02 April 2007}

  \abstract{}{We present the analysis and results of a 20 ks XMM-Newton 
            observation of RBS1423.}
            {X-ray spectral analysis is used to establish a significantly
            broadened 
            relativistic iron K$\alpha$ line from a highly ionised disk.}
            {A QSO at $z=2.262$ was considered to be the optical 
            counterpart of this ROSAT Bright Survey X-ray source.
            {Based on the improved XMM-Newton source position
            we identified a $z=0.208$ QSO as  optical 
            counterpart to RBS1423.}
            The 0.2-12 keV X-ray luminosity of this radio-quiet QSO is
            $6\times 10^{44}\,{\rm erg/s}$.
            The XMM-EPIC spectra are well described by a  power law with a
            significantly  broadened iron K$\alpha$ line.
            Disk line models for both Schwarzschild and Kerr black holes
            require hydrogen-like iron ions to fit the measured line profile.
            Significant ionisation of the reflection disk is confirmed
            by model fits with ionised disk models, resulting in
            an ionisation parameter $\xi \sim 2000$.}{}


   \maketitle
%

\section{Introduction}

The X-ray spectra of Seyfert type I galaxies and radio quiet 
quasi stellar objects (QSOs)
are generally dominated by a power law component which 
is often accompanied by a softer component below $\sim$1 keV
(``soft excess'', \citealt{arnaud}). 
An iron $\rm{K}\alpha$ fluorescence line at 6.4 keV was first observed
in  {\em OSO-8} spectra of \object{Centaurus A} (\citealt{mushotzky}). It was found to be a common
feature in the X-ray spectra of AGN by the GINGA satellite
(\citealt{nandra_pounds}).  
With the higher spectral resolution of the ASCA satellite, very broad emission features
were discovered in the X-ray spectra of a few nearby Seyfert galaxies
(e.g. \object{MCG-6-30-15}, \citealt{tanaka}; \object{NGC 3783}, \citealt{george}).
These features are usually interpreted as iron K$\alpha$ fluorescence
originating from reflection by the innermost parts of the AGN accretion
disk, where they are broadened by gravitational redshift and
relativistic motion.
With the launch of XMM-Newton and CHANDRA higher quality spectra
became available, which in many cases confirmed the relativistic
reflection scenario (e.g. for  MCG-6-30-15, \citealt{wilms},
\object{NGC 3516}, \citealt{turner02}).
While the relativistically broadened iron lines are now
found in large fraction among the local, low luminosity Seyfert galaxies 
(\citealt{nandra06}, \citealt{guainazzi}),
 detections in higher luminosity Seyfert galaxies and QSOs e.g., \object{E1821+643}
(\citealt{jimenez2007}), \object{3C109} (\citealt{miniutti}), or \object{Q0056-363} 
(\citealt{porquet}) are still rare. 
The low number of detections in luminous QSOs is partly 
due the small number of bright, low redshift, luminous QSOs.
On the other hand, it has been shown that the equivalent width of narrow 
iron K$\alpha$ emission lines is anti-correlated with X-ray luminosity
 (\citealt{iwasawa}). This so-called X-ray Baldwin effect has also been
found for the equivalent widths of relativistically broadened
iron lines (\citealt{nandra97}).
In a sample of 38 PG QSOs and (radio-loud) quasars, only three XMM-Newton spectra
showed significant detections of relativistically broadened iron lines (\citealt{jimenez}).

Here we present a 20 ksec XMM-Newton observation of the
ROSAT X-ray source RBS1423 (RA=14$^{\rm h}$\,44$^{\rm m}$\,14$^{\rm s}$, DEC=06$^{\circ}$\,32$^{\rm m}$\,30$^{\rm s}$).  
In the ROSAT bright survey (RBS, \citealt{schwope})
this source was identified with a QSO at $z=2.262$.
Based on the XMM-Newton position, we
assign a new optical counterpart to RBS1423,
a QSO with $z=0.208$ and an \mbox{0.2-12 keV} X-ray luminosity
${L_{\rm X}} = 5.7 \times 10^{44}\,{\rm erg/s}$.

The paper is organised as follows. In Sect.~\ref{section2},
Sect.~\ref{section3}, and Sect.~\ref{section21} we describe
the XMM-Newton data, the optical counterpart identification to RBS1423,
and optical observations. In Sect.~\ref{section4} we analyse the X-ray
data by fitting different models.
Finally, our conclusions are discussed in Sect.~\ref{section6}.


\section{XMM-Newton observation\label{section2}}

RBS1423 was observed by XMM-Newton on February 11, 2005 
(ObsID 0207130401, orbit 948, exposure time $\sim$20.2 ksec) with
the European Photon Imaging Cameras (EPIC) as primary instruments.
The EPIC-PN camera was operated in standard Full Window mode with
thin filter, while both EPIC-MOS cameras observed in Full Window mode
with medium filters. The data were processed with {\tt SAS} version 7.0
(Science Analysis Sofware) package, including the corresponding calibration
files. The {\tt epchain} and {\tt emchain} tasks were used for generating
linearised event lists from the raw PN and MOS data.

After cleaning the data of observation time with high background,
we obtained net exposure times of $\sim$14.1 ksec for the EPIC-PN camera and
$\sim$16.7 ksec for both EPIC-MOS cameras.
The count rates
were $CR_{\rm{MOS}}=0.392\pm0.004$ and \mbox{$CR_{\rm{PN}}=1.36\pm0.01$} per camera, respectively.
Therefore, the effects of photon pile-up are negligible for both cameras.

\section{The optical counterpart of RBS1423\label{section3}}

\begin{figure*}
   \centering
\begin{minipage}{\linewidth}
\mbox{
\begin{minipage}{6.5cm}
   \vspace*{0.8cm}\includegraphics[width=6.5cm,clip=]{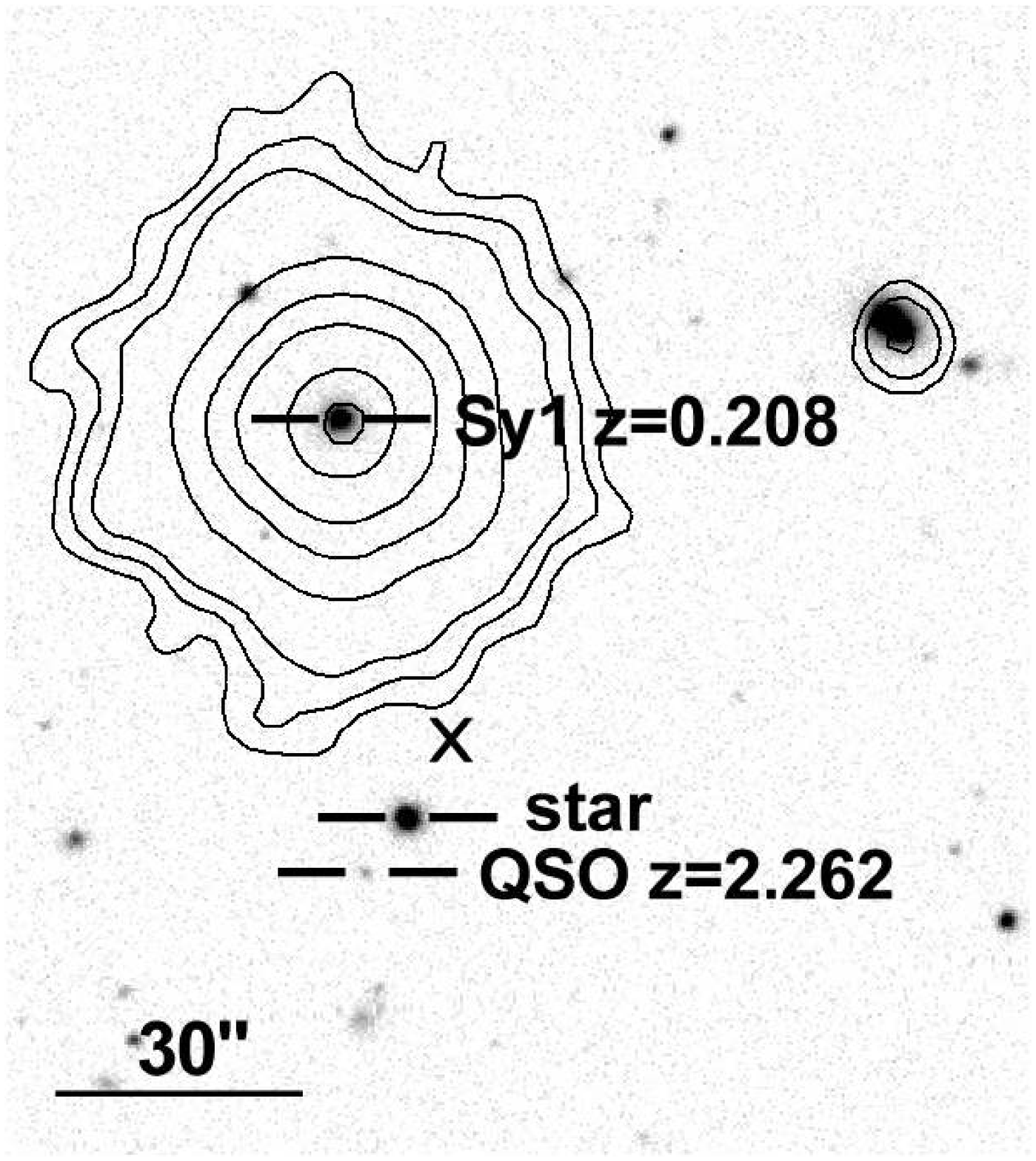}
\end{minipage}
\hspace{0.5cm}
\begin{minipage}{8cm}
   \includegraphics[width=8cm,clip=,angle=-90]{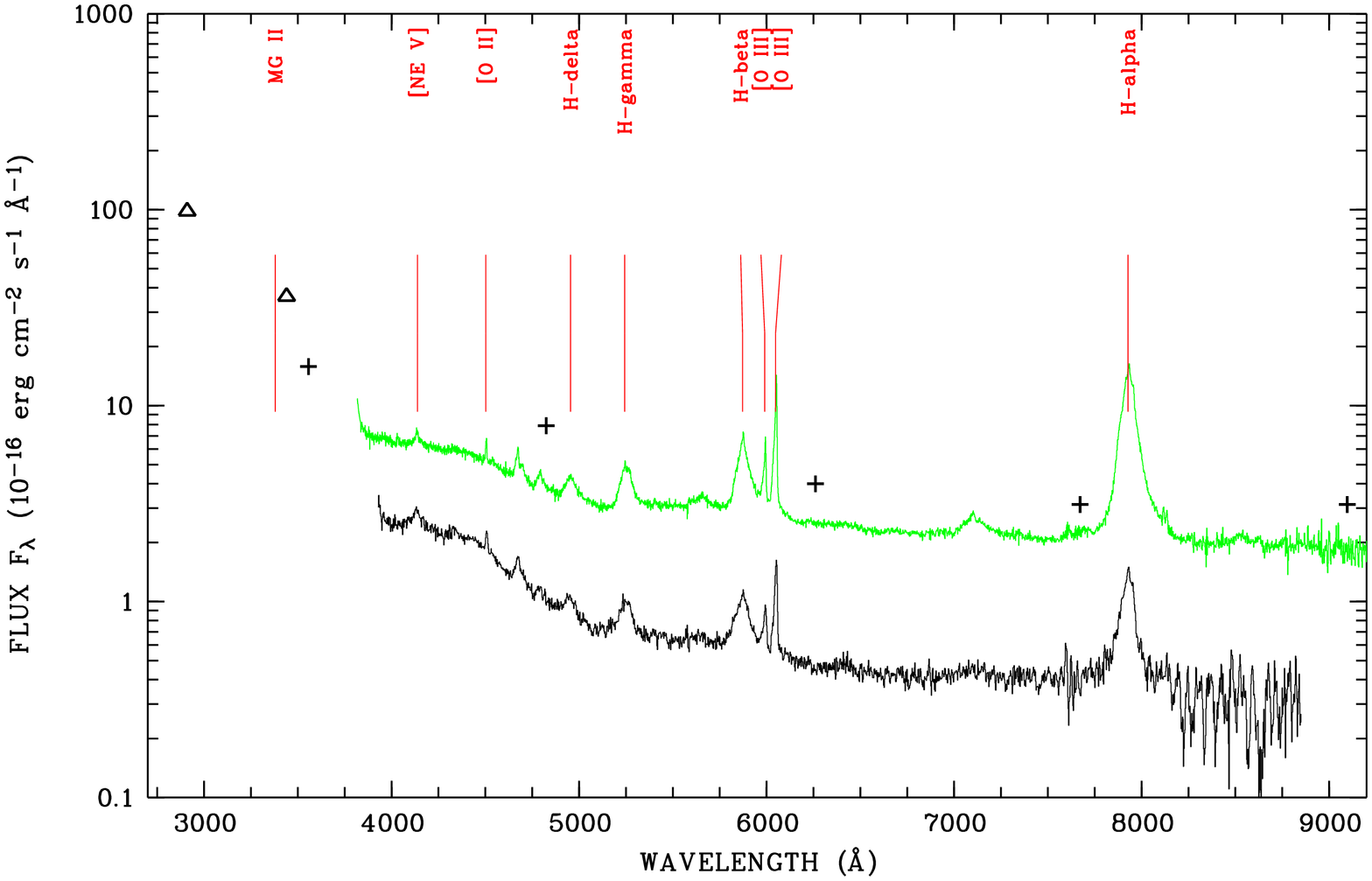}
\end{minipage}
}
\end{minipage}
\vspace*{-0.8cm}
   \caption{\emph{left:} 
            SDSS $R$-Band image of RBS1423 with overplotted XMM-Newton EPIC-PN X-ray contours. The former optical counterpart of RBS1423 (QSO
            at $z=2.262$) is clearly not associated with the XMM-Newton
            source. Instead, the QSO at $z=0.208$ (marked as Sy1) is the related
            optical counterpart. The cross indicates the ROSAT X-ray position of
            RBS1423. The X-ray source in the upper right corner is unrelated
            to RBS1423.
            \emph{right:} Optical spectrum of the
            QSO at $z=0.208$ (optical counterpart to RBS1423).
            The SDSS spectrum (green solid line) shows a significantly
            increased flux compared to the Keck spectrum (black solid line).
             XMM-Newton Optical Monitor magnitudes are plotted as triangles,
            while SDSS u-, g-, r-, i-, z-magitudes are plotted as crosses.
            The most prominent quasar emission lines are labelled.}
            \label{opt_counterpart}
\end{figure*}

The RBS-catalogue originally identified a QSO at $z=2.262$
(\citealt{schwope})
as X-ray source of RBS1423, but mentioned a possible contribution of 
a Seyfert I galaxy at 54 arcsec distance ($z=0.208$). 
Considering the measured 0.5-2.0 keV ROSAT flux
of $1.4 \times 10^{-12}$ \cgs, RBS1423 would have been one of the most 
luminous X-ray sources in the sky with $L_{\rm{X}}=5.5 \times 10^{46}\
\rm{erg\ s}^{-1}$ .

Based on the linearised event list, we produced 0.2-10 keV broad band X-ray
images for all EPIC cameras. A combined source detection was performed
({\tt edetect\_chain}) with a minimum detection likelihood $ml=5$. The source
detection list was astrometrically corrected for a systematic XMM-Newton
position error ({\tt eposcorr}). Furthermore, we generated an
astrometrically corrected EPIC-PN X-ray image, which was used to plot the
X-ray contours in Fig.~\ref{opt_counterpart}. Our XMM-Newton X-ray image
clearly reveals the Seyfert I galaxy at $z=0.208$ as the only counterpart to
RBS1423. According to the conventional dividing line of $M_{\rm B}=-23$
between Seyfert galaxies and QSOs, we classify the $M_{\rm B}=-23.0$ 
(see Sect. \ref{sect:optvar}) object as a low luminosity QSO.

The QSO at redshift  $z=2.262$ was not
detected as an X-ray source above the detection likelihood limit.
The offset between the XMM-Newton and the ROSAT position
(the latter indicated by a cross in Fig.~\ref{opt_counterpart}) is 41 arcsec.
The angular distance between the $z=2.26$ QSO and the 
$z=0.2$ QSO is 54 arcsec. Hence, the off-axis
location of the X-ray source is negligible for the further data analysis.
In total, we collected $\sim$40000 X-ray photons from RBS1423 allowing proper
spectral analysis (Sect.~\ref{s:xs}).

\section{Optical observations\label{section21}}

\subsection {Keck II \& SDSS data}
During the optical identification program of the RBS sources,
the QSO at $z=0.208$ was observed spectroscopically with
the Low Resolution Imaging Spectrograph (LRIS) at the Keck II telescope
on April 3, 1997. The longslit observation with a slit width of 1
arcsec and slit length of 175 arcsec had an exposure time of 900 seconds.
The reciprocal dispersion was 2.55 $\AA$/pixel.
The spectrum (see Fig.~\ref{opt_counterpart}, right panel, black line)
was extracted in the standard manner and standard flux calibration was
applied.
The FWHM of the H-$\alpha$ line in the Keck spectrum is $\sim$3000\,km/s
(FWHM$_{\rm{[OIII]}}\sim$700\,km/s, FWHM$_{\rm{H-\beta}}\sim$4000\,km/s).

More than four years later, on June 15, 2001, the Sloan Digital Sky Survey
(SDSS, see e.g. \citealt{adelman}) obtained photometric magnitudes in the u-, g-, r-, i-, 
z-SDSS-bands for the QSO at $z=0.2$. On May 3, 2005, an SDSS spectrum was
taken with an exposure time of 3600\,s. The SDSS spectrum is shown in
Fig~\ref{opt_counterpart}, right panel as a green line. The line widths are 
FWHM$_{\rm H-\alpha}\sim$3600\,km/s, FWHM$_{\rm [OIII]}\sim$600\,km/s, and
FWHM$_{\rm{H-\beta}}\sim$3100\,km/s.

\subsection {XMM-Newton Optical Monitor photometry}

Optical and near-ultraviolet images of RBS1423 were obtained with the optical
monitor (OM, \citealt{mason}) onboard XMM-Newton through U and UVW1 filters with central wavelengths
3440 $\AA$ and 2910 $\AA$, respectively. The exposure time in the U-filter was 5000 seconds 
and 3174 seconds in the UVW1.
The data were processed with the OM reduction pipeline {\tt omichain}.

The resulting count rates were corrected for detector dead time
and for the time-dependent degradation factor, as computed by the
{\tt SAS} task {\tt ommag}. Zero point and flux conversion factors
from the XMM-Newton calibration files allowed us to calculate optical fluxes.
The U- and UVW1-fluxes of the OM are marked in Fig.~\ref{opt_counterpart}
(right panel) as triangles.

\subsection {Optical variablility of RBS1423}
\label{sect:optvar}

We converted the SDSS $g$- and $r$-magnitudes to Johnson \mbox{$B$-magnitudes}
following \cite{smith} also applying a \mbox{$k$-correction}.
Based on the derived absolute $B$-band magnitude of $M_{\rm{B}}=-23.0$, 
we re-classify RBS1423 as a low luminosity QSO.

Comparing the different optical data sets of this QSO
(Fig.~\ref{opt_counterpart}), it is obvious that
RBS1423 is highly variable in the optical.
The SDSS spectrum shows a fivefold increased continuum flux and
a tenfold increased H-$\alpha$ line flux compared to the Keck spectrum.
The OM photometric data show a pronounced rise of the spectral flux at short
wavelength.

We searched for X-ray variability in our XMM-Newton data, but found no
significant deviation from the mean count rate.
The measured XMM-Newton flux of \mbox{$f_{\rm{X,0.5-2\,keV}}= 1.6 \times
10^{-12}$ \cgs} is comparable to the ROSAT flux.
Multiple observations covering longer time scales are required to observe a possible
X-ray variability.

\subsection {Black hole mass estimate\label{blackhole}}
\begin{figure}
   \centering
\resizebox{\hsize}{!}{ 
   \includegraphics[bbllx=24,bblly=17,bburx=346,bbury=272]{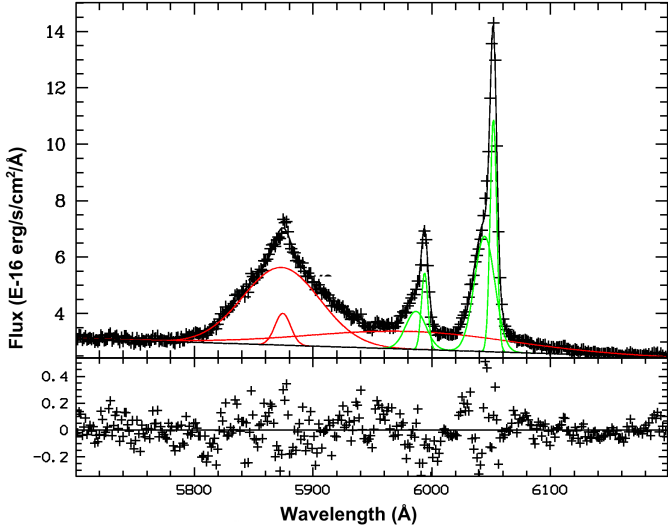}} 
      \caption{Line profile analysis of H-$\beta$ and O{\sc [iii]} system.
               The upper plot shows the measured SDSS spectrum of RBS1423 
               in black crosses.
               Different line model components are plotted in green 
               (for O{\sc [iii]}) and red solid lines (H-$\beta$). 
                The lower panel
               indicates the residuals between summed line models and 
               observed data (same units).}
         \label{h-beta}
\end{figure}

Since the optical spectra cover the H-$\beta$ line, we are able to 
estimate the black hole mass of RBS1423 following \cite{vestergaard}. 
The analysis of the H-$\beta$ and O{\sc [iii]} system (Fig.~\ref{h-beta}) 
reveals a superposition of a narrow-line component and a blueshifted 
broad-line component. The SDSS spectrum was normalised to the photometric 
$r$-magnitude to compensate for slit losses during spectroscopy. 
This is essential, since the optical continuum luminosity 
L$_{\lambda,{\rm rest}}(5100 \AA$) is needed to estimate the black hole mass.
The observed FWHM of the summed H-$\beta$ line components is 61\,$\AA$. We
  correct the line width measurement for spectral resolution following 
 \cite{peterson} and take into account the SDSS resolution of R$\sim$2000. 
Based on a computed $\lambda$\,L$_{\lambda,{\rm rest}}(5100 \AA$) $\sim$
  $3.2 \times 10^{44}$\,erg/s, the black hole mass is estimated to 
log\,$M_{\rm BH}$=8.1.

\section{X-ray spectral analysis\label{section4}}
\label{s:xs}

X-ray spectra of RBS1423 were extracted from calibrated photon event lists
with {\tt SAS} task {\tt especget} which also
creates the appropriate redistribution matrices and exposure-corrected
effective area files.
Instead of using the published effective area file shipped with {\tt SAS} 7.0
for EPIC-PN we used the more recent version 
{\tt XRT3$\_$XAREAEF$\_$0011.CCF} kindly made available by F.~Haberl (MPE).
This version implements a slight reduction of the XRT3 effective areas 
(with respect to XRT1 and XRT2) in the 6-8 keV range, 
which had been measured during the pre-flight calibration of the telescopes. 
The new version improves the concordance of the MOS and PN spectra 
at the energy of the Fe $K_\alpha$ line. 

X-ray events corresponding to patterns 0-12 and 0-4 events (single and double
pixels) were selected for MOS and PN. For the EPIC-PN camera
only X-ray events with flag 0 were included.  The MOS X-ray events were
extracted by using the recommended flag selection {\tt XMMEA\_SM}. A low-energy
cutoff was set to 0.2 keV, while the high-energy cutoff was fixed to 12.0 keV.
Identical source regions were used for all EPIC cameras in order to extract
spectra. All spectra were binned to a minimum of 30 counts per bin to apply
the $\chi^2$ minimisation technique in the spectral fitting, which was
perfomed by using the {\tt XSPEC} package version 12.0.
All quoted errors are 68\% limits unless otherwise mentioned.

The galactic hydrogen column density in the line of sight is
$N_{\rm H_{\rm{GAL}}}=2.72 \times 10^{20}\,{\rm cm^{-2}}$ (\citealt{dickey}). 

\begin{figure}
   \centering
 \includegraphics[width=6cm,angle=-90,clip=]{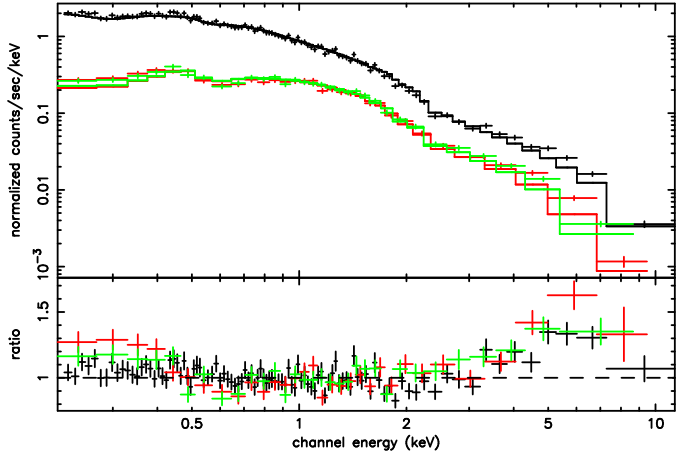}
      \caption{PN (black) and MOS (green, red) cameras spectra of RBS1423 (in observer frame).
              A single power law has been fitted to the \mbox{0.2-12.0 keV} data. A broad
              and significant positive residual is seen between \mbox{$\sim4-7$ keV,}
              suggesting the presence of an iron K$\alpha$ line. In the soft
              energies (0.2-0.5 keV) a discrepancy between the XMM-Newton
              camera types and a possible soft excess is noticed.
              The data have been rebinned to
              a signal-to-noise ratio (SNR) of 15, after grouping to a
              minimum of 30 counts per bin.
              }
         \label{xrayspec}
\end{figure}

A single power law model with galactic absorption was 
fitted to the broad-band 0.2-12.0 keV PN and MOS spectra. The best fitting
power law ($\chi^2$/d.o.f.$=1051/754$ with $\Gamma=2.05\pm0.01$)
showed significant positive residuals around $\sim$5 keV, typical for iron
K$\alpha$ emission (Fig.~\ref{xrayspec}). 
Below 0.5 keV, a discrepancy between PN and both MOS
cameras is noticable ($\sim$15\% higher count rates in MOS cameras relative to PN).
The PN data are broadly consistent with a power law emission spectrum and
foreground galactic absorption,
while the MOS data indicate a possible soft excess.
Due to this discrepancy we use only the spectra above \mbox{0.5 keV} for the analysis
of the possible  Fe-K$\alpha$ feature.
Moreover, we fit the PN and MOS data separately.

All fit results are summarised in
Table~\ref{pnresults} for the PN detector and in Table~\ref{mosresults} for
the MOS cameras. Line energies and equivalent widths presented in these
tables are given in the source frame.

\begin{figure}
   \centering
 \includegraphics[width=6cm,angle=-90,clip=]{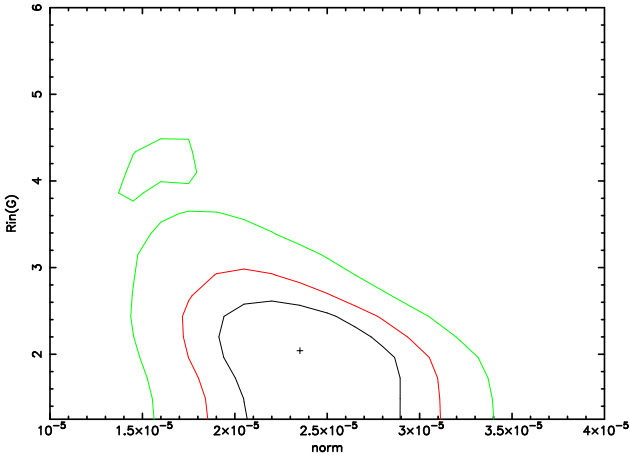}
      \caption{Confidence contours (68\%, 90\%, 99\%) of the
               parameters $R_i$ and line normalisation for
               a single line with {\tt laor} profile. Both PN and
               MOS spectra have been used to calculate the contours.
              }
         \label{laor_contour}
\end{figure}

The line feature is also apparent, when we fit single power laws 
to the PN and MOS data separately in the energy range \mbox{0.5-12.0 keV.}
We then tried to fit the spectra by introducing a second
power law (labelled as {\tt wabs(power1 + power2)}). However,
the remaining excess still indicates the presence of a
broad emission line feature.

As the next simple approach, we tried to fit the data using
a single power law with a Gaussian line superposed. 
The line was detected at $(4.68\pm0.43)$ keV with a $\sigma=(1.78\pm0.41$) keV
in the PN detector and at $(5.23\pm0.19)$ keV ($\sigma=0.86\pm0.23$ keV) in the
MOS detectors. This corresponds to a rest-frame energy of \mbox{5.6 keV} and 6.3 keV,
respectively. Considering the statistical errors, the line energies
measured by the MOS and PN cameras differ only marginally. However, some
systematic influence of the effective area calibrations on the line energies
cannot be ruled out. 

The significant width of the line suggests
relativistic broadening of the line.
We therefore fitted iron K$\alpha$ disk line
models to the spectra: The {\tt XSPEC} model {\tt diskline} (\citealt{fabian89})
calculates the line broadening in an accretion disk around a non-rotating 
(Schwarzschild) black hole (BH)
and the model {\tt laor} (\citealt{laor}) around a spinning (Kerr) BH.  
The line emissivity is
parameterised by a power law as a function of radius, $R^{-q}$ with $q$ 
as a free parameter. 
Since $q$ is unconstrained we used an accretion disk
emissivity law $(1-\sqrt{6/R})/R^3$ for the Schwarzschild BH model 
and $q=3$ for the Kerr BH model ($R^{-3}$).

The relevant model parameters and the equivalent widths of the emission line
fits are given in Tables \ref{pnresults} and \ref{mosresults}. 
For both cameras the {\tt laor} fit results are statistically the better
fits, favouring the rotating black hole scenario.
Fig.~\ref{laor_contour} shows that for the {\tt laor} model 
the emission line is required with high significance and that it
is significantly broadened in the relativistic regime. 
The rest-frame line energy for a simultaneous fit of the PN and MOS data is 
(7.45 $\pm$ 0.07) keV.
Even if a broken power law plus a relativistic disk model {\tt laor} 
is used, the broad-line feature is dectected with a significance of more 
than 3$\sigma$ in the PN and MOS data.

In none of the cases is the source frame energy of the line consistent with
the K$\alpha$ energy expected for neutral iron \mbox{(6.4 keV)}.
Instead, the fits suggest that the iron feature is due to fluorescence
from hydrogen-like iron with a K$\alpha$ energy of \mbox{6.9 keV.}
Fixing the source frame line energy to 6.4 keV and fitting the 
accretion disk inclination angle in the model {\tt laor} as a 
free parameter formally leads to acceptable $\chi^2$ values, but leaves
visible residuals in the range beyond 6.4 keV.

\begin{table*}
\caption{Results of the spectral analysis for the PN detector (0.5-12.0 keV). Notation of the
  models is according to {\tt XSPEC}. Frozen parameters are between exclamation
  marks. Energies and equivalent width are given in the quasar frame.}
\label{pnresults}
\centering
\begin{tabular}{c c c c c c c c c} 
\hline\hline
Model & $\Gamma$ &  $E_{\rm{Fe,\, rest}}$& $EW_{\rm{rest}}$ & $R_{\rm{IN}}$ & $R_{\rm{OUT}}$ & $i$ & $\chi^2$/d.o.f.\\
  &  & keV & keV & r$_{\rm {Schwarzschild}}$ & r$_{\rm {Schwarzschild}}$ & deg & \\
\hline
   wabs(power) & 2.04$\pm$0.01  & - & - & - & - & - &
   477/378\\
   wabs(power1 + power2) & 1.58$\pm$0.30 \& 2.51$\pm$0.35 &- & - & -
   & - & - &436/376\\
   wabs(power + gauss) & 2.14$\pm$0.03&5.65$\pm$0.52 & $2.30^{+1.67}_{-0.69}$  & - &   - & - & 418/375\\
   wabs(power + diskline)
  &2.06$\pm$0.01&7.10$\pm$0.08&$0.63^{+0.19}_{-0.21}$&6$\pm$110&!400!&!30!&455/375\\
   wabs(power + laor) &2.08$\pm$0.02&7.47$\pm$0.09&$1.60^{+0.35}_{-0.41}$
   &1.9$\pm$2.7&!400!&!30!&440/375\\
   wabs(po1 + po2 + laor) &1.55$\pm$0.54 \& 2.36$\pm$0.36&7.10$\pm$0.13&$0.77^{+0.33}_{-0.32}$
   &!1.9!&!400!&!30!&424/374\\
   wabs(kdblur(power + reflion)) & 1.62$\pm$0.24 & - & - & 2.8$\pm$1.6 & !400! & !30!&  421/374\\
\hline
  & & \multicolumn{2}{c}{cov. fraction}& \multicolumn{2}{c}{$N_{\rm{H}}$ in 10$^{22}$ cm$^{-2}$}&  & \\ 
wabs(zpcfabs(power))& 2.14$\pm$0.02& \multicolumn{2}{c}{0.38$\pm$0.04}& \multicolumn{2}{c}{17$\pm$4}&  &  420/376\\ 
\hline                 
\end{tabular}
\end{table*}

\begin{table*}
\caption{Results of the spectral analysis for the MOS detectors (0.5-12.0 keV). Notation of the
  models is according to {\tt XSPEC}. Frozen parameters are between exclamation
  marks. Energies and equivalent width are given in the quasar frame. }
\label{mosresults}
\centering
\begin{tabular}{c c c c c c c c c}
\hline\hline
Model & $\Gamma$ & $E_{\rm{Fe,\,rest}}$& $EW_{\rm{rest}}$&  $R_{\rm{IN}}$ & $R_{\rm{OUT}}$ & $i$ & $\chi^2$/d.o.f.\\ 
  &  & keV & keV & r$_{\rm {Schwarzschild}}$ & r$_{\rm {Schwarzschild}}$ & deg & \\
\hline
   wabs(power) & 1.90$\pm$0.02  & - & - & - & - & - &
   314/280\\
   wabs(power1 + power2) & 1.25$\pm$0.83 \& 2.14$\pm$0.32 &- & - & -
   & - & - &301/278\\
   wabs(power + gauss) & 1.96$\pm$0.02& 6.32$\pm$0.23 & $1.00^{+0.30}_{-0.27}$  & - &   - & - & 286/277\\
   wabs(power + diskline)
   &1.93$\pm$0.02&7.42$\pm$0.12&$0.59^{+0.25}_{-0.25}$&6$\pm$571&!400!&!30!&302/277\\
   wabs(power + laor) &1.96$\pm$0.02&7.49$\pm$0.15&$1.28^{+0.42}_{-0.39}$
   &2.4$\pm$1.6&!400!&!30!&291/277\\
   wabs(po1 + po2 + laor) &!1.25! \& 2.04$\pm$0.07&6.85$\pm$0.15&$0.96^{+0.39}_{-0.39}$
   &!2.4!&!400!&!30!&287/277\\
   wabs(kdblur(power + reflion)) & 1.72$\pm$0.16&-&-& 60$\pm$600 & !400! & !30! & 285/276\\
\hline
 & & \multicolumn{2}{c}{cov. fraction}& \multicolumn{2}{c}{$N_{\rm{H}}$ in 10$^{22}$ cm$^{-2}$}&  & \\ 
wabs(zpcfabs(power))& 1.97$\pm$0.02& \multicolumn{2}{c}{0.37$\pm$0.09}& \multicolumn{2}{c}{38$\pm$16}&  &  293/278\\ 
\hline
\end{tabular}
\end{table*}

Since broad iron K$\alpha$ features in AGN are connected with 
reflection by the accretion disk, other effects modifying
the reflected X-ray spectrum have to be taken into account.
We therefore applied the self-consistent ionised disk reflection models
by \cite{ross} to model the X-ray spectrum of RBS1423.
Their calculations include a range of relevant fluorescence lines, absorption
edges, Compton-reflection, and Compton-broadening of the atomic features.
The models are available in tabulated form for fitting with {\tt XSPEC} ({\tt reflion})
and can be convolved with relativistic blurring using the code by
\citealt{laor} (model {\tt kdblur}).

The observed X-ray spectrum is then interpreted as superposition of the illuminated
power law and the reflected component, described by the {\tt XSPEC} models
{\tt wabs(kdblur(power+atab$\{$reflion.mod$\}$))}.

\begin{figure}
   \centering
 \includegraphics[width=6cm,angle=-90,clip=]{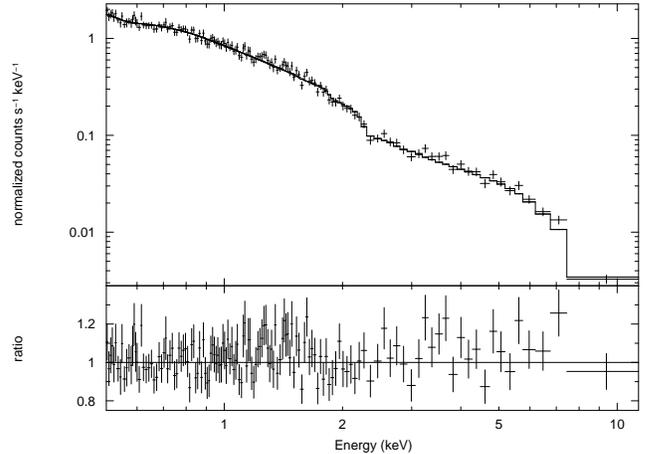}
      \caption{PN spectra of RBS1423 (observer frame).
              A relativistically blurred self-consitent ionised disk
              reflection model ({\tt kdblur, reflion}) has
              been fitted to the 0.5-12.0 keV data. The comparison with
              Fig.~\ref{xrayspec} shows that the broad positive deviation in
              the $\sim4-7$ keV region are well fitted. For
              presentation purpose, the data have been rebinned into groups of bins
              with a signal-to-noise ratio (SNR) of 10, after grouping to a
              minimum of 30 counts per bin.
              }
         \label{xrayspec2}
\end{figure}

Since the data quality is not sufficient to constrain all parameters, we set the
outer radius of the relativistically blurring model to 400 Schwarzschild
radii and the inclination to 30 degrees. This choice is not critical to our results.
Furthermore, the power law of the
reflected component is chosen to be equal to the power law of the illuminating
radiation. Again, we fitted the PN and MOS data separately. Within the
calculated errors, the fit parameters agreed very well for both data sets.
The $\chi^2$ values of the {\tt reflion} fits (see. Fig.~\ref{xrayspec2}) 
are slightly better
than those of the {\tt laor} and {\tt diskline} fits.
The iron abundance was found to be nearly solar (Fe$_{\rm{PN}}$= 0.8$\pm$0.2, 
Fe$_{\rm{MOS}}$= 1.2$\pm$0.9). 
This value seems to be contradictory to the large equivalent widths 
of the iron line resulting from the fits including a power law and 
a gaussian or relativistically blurred line.
Indeed, if we modify the best fit {\tt reflion} model by setting the 
iron abundance to the lowest possible value (A=0.11) and replace the
iron emission feature by a Gaussian line, we get an estimate of 170 eV for the
equivalent width of the iron line.    
It is likely that the iron line fluxes from the power law 
based fits are overestimated, since these models do not account for
the effect of the iron K absorption edge and as a result the slope of the power law
becomes too steep. 
This explanation is confirmed by the fact that the iron 
equivalent width is largest for the fits with the steepest
power law models (see Table \ref{pnresults} and \ref{mosresults}).
Therefore we consider the results of the physically  consistent {\tt reflion}
model as more reliable.

\begin{figure}
   \centering
 \includegraphics[width=6.5cm,angle=-90,clip=]{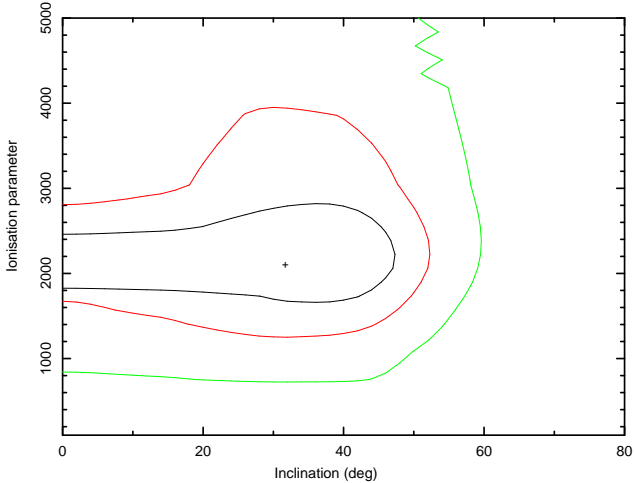}
      \caption{PN confidence level contours (68, 90, and 99 per cent) 
              for the relativistically blurred disk reflection. The ionization
              parameter is plotted in dependence on the accretion disk
              inclination angle. 
              }
         \label{contour}
\end{figure}

The best fitting ionisation parameters are
$\xi_{\rm{PN}}$=2000$\pm$500 and $\xi_{\rm{MOS}}$=2000$\pm$800.
This result confirms that the reflecting parts of the accretion disk in
RBS1423 are highly ionised. Figure~\ref{contour} shows that the data
cannot be fitted with a lower ionization parameter value and a different
accretion disk inclination angle. Independent of the inclination angle, a 
highly ionised disk is needed.

\begin{figure}
   \centering
 \includegraphics[width=6cm,angle=-90,clip=]{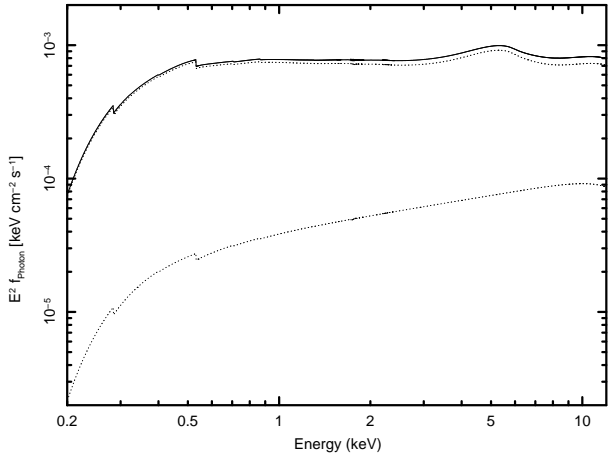}
      \caption{E$^2$\,f(E) - model spectrum for the RBS1423 PN data with a
               {\tt wabs(kdblur(power + reflion))}-fit. The lower dotted line
              represents the illuminating power law radiation, while the upper
              dotted line shows the reflected component by the accretion
              disk. Both components were gravitationally blurred and absorbed
              by the galactic hydrogen column density. The solid line
              indicates the superposition of both components. Note that the
              normalisations of the direct and reflected components are highly degenerate.
              }
         \label{model}
\end{figure}

Fig.~\ref{model} shows the E$^2$\,f(E)-model spectrum for the {\tt reflion} fit
to the PN data. The best fit model is reflection-dominated with a very weak direct power law
component. However, the errors in the normalisations of both the direct and reflected components are
rather large.

The 0.2-12.0 keV X-ray flux of RBS1423 according to the model is 
$f_{\rm{X}}\simeq 4.7 \times 10^{-12}$\,erg\ cm$^{-2}$\ s$^{-1}$. The
rest-frame X-ray luminosity (0.2-12.0 keV) is
$L_{\rm{X}}\simeq 5.7 \times 10^{44} \mathrm{\,erg~s^{-1}}$.
This makes RBS1423 one of the few X-ray luminous AGN with a detected
broad iron fluorescence line. 

Assuming $L_{\rm{X}}$ is representing the
bolometric luminosity, the accretion rate $\dot{M}$ can be calculated by using  
\begin{equation}
L_{\rm{X}} = L_{\rm{bol}} = \frac{G M_{\rm BH} \dot{M}}{R_{\rm BH}} =\nu \dot{M} c^2\ \ ,
\end{equation}
where $G$ is the gravitational constant and $\nu$ the efficiency parameter. If
we set $\nu$=0.1 the accretion rate is $\dot{M}=0.1 \rm{M}_{\odot}$/yr. The further
assumption that the accretion rate is 10\% of the Eddington luminosity, leads
to an approximate black hole mass of log\,$M_{\rm BH}$=7.6. The discrepancy
between optically and X-ray determined black hole mass of a factor $\sim$3 
is consistent with the uncertainties of the used assumptions.

As an alternative to the disk refelection models we fitted a partial covering model 
({\tt XSPEC} model {\tt zpcfabs}) to the
EPIC spectra (see Tables \ref{pnresults} and \ref{mosresults}) . 
The best fit absorbing column density is $N_{\rm H}$$\sim$$2 \times
10^{23}$\,cm$^{-2}$ with a covering fraction of $\sim$40\% (PN). 
With $\chi^2$ values of $\chi^2$/d.o.f.=420/376 (PN) and
$\chi^2$/d.o.f.=293/278 (MOS) the partial covering model fits the data equally as
well as the ionised reflection model.
Since the X-ray spectrum of RBS1423 does not show a strong
soft excess, models with partial coverage by an ionised absorber do not fit the data
well. For the PN spectrum the 90\% upper limit for the ionisation parameter is
$\xi=115$.

\section{Discussion and conclusions\label{section6}}

We analysed an XMM-Newton observation (PN net time exposure
$\sim$14.1 ksec) of RBS1423, a suspected high-luminosity QSO. 
We showed that the corresponding optical
counterpart is a broad-line AGN at $z=0.208$, thus correcting the 
identification in \cite{schwope}.
The absolute magnitude in the $B$-band of $M_{\rm{B}}\sim-23.0$ and the X-ray
luminosity of $L_{\rm{X}}\simeq 5.7 \times 10^{44} \mathrm{\,erg~s^{-1}}$
(0.5-12 keV) classify the objects as a  QSO.
Several optical observations clearly indicate  high optical
variability and a strong UV excess. 

In the X-ray spectrum of RBS1423 we detected
a broad emission feature around $\sim$5 keV,
which is well described by iron K$\alpha$ fluorescence
in the innermost parts of the accretion disk.
This makes RBS1423 one of the most luminous radio-quiet AGN
with a detection of relativistic disk reflection.

Alternatively, the X-ray spectrum can be modelled equally well 
with a partially covered power law model.
The required column density is $N_{\rm H}$$\sim$$2 \times 10^{23} {\rm cm^{-2}}$
with a  covering fraction of $40\%$.
Since the shape of a highly absorbed component is similar to the shape of
a relativistically broadened iron line, even much better signal-to-noise  
EPIC spectra of Seyfert galaxies can be explained by both reflection 
and partial covering models.
\cite{fabian04} successfully fitted the variable XMM-Newton 
spectra of the Seyfert galaxy \object{1H 0707-495}  
with ionised reflection models, while \cite{gallo}  modelled the
same data with partial absorption.

The spectrum of \object{RBS1423} does not show a strong 
soft excess. Therefore, the partial absorber must be neutral to avoid 
a strong contribution to the spectrum below 1 keV.

For a rather luminous AGN like \object{RBS1423} a two-component 
model, where one component is not absorbed at all and the second 
is absorbed by a dense column of neutral gas, is hardly feasible
(see also the discussion in \cite{miniutti} on the possibility of a partial absorber
in \object{3C 109}).
Therefore, we clearly favour disk reflection models to interpret 
the RBS1423 spectrum.

The energy of the line resulting from disk line fits is not
consistent with K$\alpha$ from neutral iron, but suggests
the presence of highly ionised iron.
This is confirmed by ionised disk model fits ({\tt reflion}, \citealt{ross})
resulting in an ionisation parameter $\xi \sim 2000$.
We compared the self-consistent ionised disk model fits with models
including a power law and a broadened emission line. 
We conclude that
the large equivalent line widths of the power law models are
likely to be overestimated, since the underlying power law tends to be
modelled too steeply when the iron edge is not taken into account.

Our detection adds a new piece of information to the ongoing debate about whether
the accretion disks in high luminosity AGN
are generally more highly ionised than in low
luminosity objects. \cite{nandra97} find evidence for a change in the
iron K$\alpha$ line profile with increasing luminosity
and also find an anti-correlation of broad-line equivalent width
and luminosity (X-ray Baldwin effect).
They attribute this correlation to the fact that ionisation
increases with luminosity.
Interestingly, the line profile in RBS1423 is similar to the
average profile of the AGN in the $10^{44}< L_X < 10^{45}$
range in \cite{nandra97}, nearly symmetric with a strong blue wing. 

On the other hand, no evidence for strong ionisation has been found
in the disk reflection spectra of the luminous quasars \object{PG 1425+267}
(\citealt{miniutti_fabian}) and 3C 109 (\citealt{miniutti}).
However, these quasars are radio-loud and therefore not
fully comparable with RBS1423. 

\citealt{porquet} found  significant broad K$\alpha$ reflection from neutral
 iron in the radio-quiet QSO \object{Q0056-363}, which is about twice as X-ray luminous
as \object{RBS1423}. They conclude that the accretion rate of the object has to be less than 
$5\%$ of the Eddington accretion rate to avoid ionisation of the disk. This is at odds 
with other estimates of the accretion rate for this object.
In another high luminosity radio-quiet QSO, \object{E1821+643},
a double peaked Fe $K_\alpha$ profile was detected in the EPIC spectrum (\citealt{jimenez2007}).
The first peak is consistent with neutral iron, the second with \ion{FE}{xxv} or
\ion{Fe}{xxvi}.
However, the ionised iron line in the object could also be due to contamination 
by emission from a surrounding cluster of galaxies.

In summary, we strongly favour the relativistic disk line reflection over the
partial covering model. The physically self-consistent ionised disk reflection
model (\citealt{ross}) yields a robust detection of disk ionisation.
Therefore, we conclude that RBS1423 is one of the first radio-quiet QSO for
which ionised disk reflection is established.

\begin{acknowledgements} 
Mirko Krumpe is supported by the Deutsches
Zentrum f\"ur Luft- und Raumfahrt (DLR) GmbH  
under contract No. FKZ 50 OR 0404.
Georg Lamer acknowledges support by the Deutsches Zentrum f\"ur Luft- und  
Raumfahrt (DLR) GmbH under contract no.~FKZ 50 OX 0201.
\end{acknowledgements}

\end{document}